\documentclass[]{aastex}
\usepackage{spr-astr-addons}

\usepackage{textcomp}
\usepackage{multirow}
\usepackage{mathtools}

\begin{document}

\title{Data Processing Pipeline for Pointing Observations of Lunar-based Ultraviolet Telescope}

\shorttitle{Data Processing Pipeline for LUT}
\shortauthors{Meng et al.}

\author{Xian-Min Meng\altaffilmark{1,2}, Li Cao\altaffilmark{1,2}, Yu-Lei Qiu\altaffilmark{1,2}, Chao Wu\altaffilmark{1,2}, Jing Wang\altaffilmark{1,2}, Xu-Hui Han\altaffilmark{1,2}, Jin-Song Deng\altaffilmark{1,2}, Li-Ping Xin\altaffilmark{1,2}, Hong-Bo Cai\altaffilmark{1,2}, Jian-Yan Wei\altaffilmark{1,2}}
\altaffiltext{1}{National Astronomical Observatories, Chinese Academy of Sciences, Beijing 100012, China}
\altaffiltext{2}{Key Laboratory of Space Astronomy and Technology, National Astronomical Observatories, Chinese Academy of Sciences}
\email{mengxm@nao.cas.cn}

\begin{abstract}
We describe the data processing pipeline developed to reduce the pointing observation data of
Lunar-based Ultraviolet Telescope (LUT), which belongs to the Chang'e-3 mission of the Chinese Lunar Exploration Program. 
The pointing observation program of LUT is dedicated to monitor variable objects in a near-ultraviolet (245--345\,nm) band.
LUT works in lunar daytime for sufficient power supply, so some special data processing strategies have been developed for the pipeline.
The procedures of the pipeline include stray light removing, astrometry, flat fielding employing superflat technique, 
source extraction and cosmic rays rejection, aperture and PSF photometry, aperture correction, and catalogues archiving, etc. 
It has been intensively tested and works smoothly with observation data.
The photometric accuracy is typically $\sim$0.02\,mag for LUT 10\,mag stars (30s exposure),
with errors come from background noises, residuals of stray light removing, and flat fielding related errors.
The accuracy degrades to be $\sim$0.2\,mag for stars of 13.5\,mag which is the 5$\sigma$ detection limit of LUT.
\end{abstract}


\section{Introduction} 
\label{sec:Intro}

Lunar-based Ultraviolet Telescope (LUT) is the first robotic astronomical telescope
deployed on the moon surface.
LUT is placed inside a cabin of the lander of Chang'e-3 mission \citep{Ip2014RAA}
which belongs to the Chinese Lunar Exploration Program. 
After the successful landing on the moon of the lander in December 2013, 
LUT has been working continuously up to the present.
It works about 12 terrestrial days per month in lunar daytime for sufficient power supply.
LUT observation has two key programs: pointing observation program and survey program.
The pointing program monitors brightness of variable targets which are proposed by world-wide astronomy
society, including cataclysmic variables, RR Lyrae stars, eclipsing binaries, 
active chromosphere stars, flaring M dwarfs, etc.
From the first light on Dec. 15, 2013 to Feb.~2015, LUT has monitored variable stars for $\sim$800 hours,
and surveyed an area of $\sim$1600\,$\deg^2$ around the moon's north pole.

Typically, one target is continuously monitored for $\sim$50 hours, which generates about two thousands of images
and requires about 4.5~Gbytes storage space.
Data processing is not performed real-time because data are obtained after LUT finished its lunar daytime work.
Data processing of pointing observations should be finished in a handful of days so as to leave enough data processing time for other observation programs.
Therefore, an automatic data processing pipeline is mandatory to address the issue of massive data quantity.
The most parts of the pipeline follow the astronomy data reduction routines, 
while some parts are developed for special features of LUT observations.
For example, the LUT images suffer from significant pollution of stray light from 
scattered sunlight in the cabin and the telescope.
Another pollution source comes from the cosmic rays in space,
which is much more significant than that on earch. 
Furthermore, flat field images taken with internal LEDs are not perfect 
in terms of large-scale uniformity. 
To solve this problem, the large-scale illumination structure,
i.e. superflat, obtained through dithering observations, is coupled with the illumination corrected and normalized LED flat field.

This paper describes the data processing pipeline reducing the pointing observation data of LUT.
A brief description of LUT's instruments and pointing observations are presented in Section~\ref{sec:Obser}.
Details of the procedures of the pipeline, and the building of LUT flat field
are described in Section~\ref{sec:Pipeline}.
Section~\ref{sec:Apertcor} describes the aperture correction method.
The precision of the photometry after the stray light removing and flat fielding 
are shown in Section~\ref{sec:Discussion}.

\section{Instrument, Observation \& Calibration}
\label{sec:Obser}

LUT is a 150\ mm, F/3.75 Ritchey-Chretien telescope working at a Nasmyth focus.
A flat mirror is mounted on a two-dimensional gimbal in front of the telescope aperture
for pointing and tracking \citep[see][Fig.~1]{Wangjing2015cali}.
The mirror can rotate from -28\textdegree\ to +13\textdegree\ in azimuth (axis of telescope as zero),
and +20\textdegree\ to +38\textdegree\ in altitude (horizontal direction as zero). 
A UV-enhanced back-illuminated AIMO CCD is mounted on the fucal plane,
and UV coating is applied on one lense of the field corrector as the UV filter.
The resulting passband of LUT is about 245--345\,nm, peaking at 250\,nm.
The CCD pixel scale is 4.76''\,pixel$^{-1}$, so the exposure area of CCD with 1024$\times$1024 pixels gives 
a field of view (FOV) 1.35\textdegree$\times$1.35\textdegree.
Two pairs of LEDs are installed crosswise (one as backup) on the front inside wall of the CCD camera, 
which can be used to illuminate the CCD through a ring-like diffusing glass for flat field calibration.
The LEDs emit at 286\,nm and the spectral widths are $\sim$12\,nm.
For further details of the scientific objectives, instrumentation, system performance, 
and calibrations of LUT, please refer to \cite{Caoli2011} and \cite{Wangjing2015cali}.

The pointing observation strategy is described as follows.
A target is placed near the center of the CCD and monitored for several observational runs.
Each run lasts for about 30 minutes, consists of several exposures,
and has a fixed telescope pointing with respect to the moon.
During a run, the total shift of the target due to the rotation of the moon 
is within a region of $\sim 50 \times 100$\,pixel.
The shift of stars in image during each exposure is small (within 1\,pixel) comparing with their profile widths of about 2\,pixels 
(see Sect.~\ref{subsec:sextractor}).
The next run re-direct the flat mirror pointing to make the target return to the center.
Such a strategy favors stray light removing and flat fielding (see Sect.~\ref{sec:Pipeline}).

\begin{figure}[!htbp]
  \centering
  \includegraphics[width=\columnwidth]{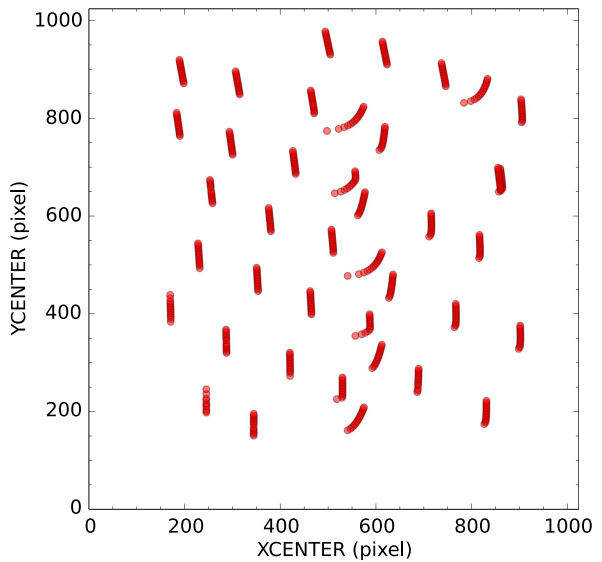}
  \caption{A typical dithering observation sampling grid of the flat field taken in June 17, 2014. 
  The target star was HD\,152303 with signal-to-noise ratio (SNR) 66.8 measured in 2$\times$FWHM aperture radius. XCENTER and YCENTER are column- and row-direction coordinates of the star on every image frame.}
  \label{fig:flatdither}
\end{figure}

Calibration observations include dark field acquisition, internal flat field exposures and superflat observations. 
LUT can obtain internal flat field images to correct pixel-to-pixel nonuniformity in CCD sensitivity,
making use of its LED lamps.
However, the LED illumination is not ideal in terms of large-scale uniformity.
To correct the large-scale nonuniformity, superflat images are created employing dithering observation technique.
Before dithering observation, a positional grid was designed to sample the large-scale nonuniformity structure in the FOV.
Usually, a grid size of 7$\times$7 was adopted (see Fig.~\ref{fig:flatdither}).
At each nodal point, the standard star was observed for about 20 times.

Superflats are not created making use of the sky background, because the atmosphere on moon is extremely tenuous so there 
is no sky background available, and also because the ecliptic light is not in LUT's available sky area.
But after all, the stay light suffered by LUT would certainly contaminate the flat fields.
Therefore, the superflat is actually uncovered using standard stars as uniform light sources.
The dithering observation was carried out at most once for each month, 
depending on both the Chang'e-3 and LUT operation plan arrangement.
They had been carried out in January, June, August and December 2014, and in January and May 2015.
Flat field correction in each month made use of the superflat of the adjacent month.

\section{Pipeline Description}
\label{sec:Pipeline}

The data processing pipeline is developed for LUT pointing observations data reduction, focusing on obtaining catalogues and light curves of targets of interest.
The pipeline consists of several procedures, including overscan correction, stray light removing, 
astrometry, flat fielding, source extraction, cosmic ray rejection,
aperture radii determination, photometry, aperture correction and catalogues archiving procedures,
which are summarized in Table~\ref{Tab:PipelineDetail}.
Each procedure is described in the following subsections. 
The pipeline is developed with {\sc SExtractor}, {\sc IRAF} and {\sc PyRAF} ({\sc IRAF} wrapped in {\sc Python}) programs, 
and {\sc Python} packages including {\sc NumPy}, {\sc SciPy}, {\sc AstroPy}, {\sc PyFITS} and {\sc astLib}.

\begin{table*}[!htbp]
\centering
\caption[]{Outline of Data Processing Pipeline}
\begin{tabular}{|l|}
\hline
{\bf Data Processing Outline}\\
\hline
0. Data Preparation\\
1. Overscan Correction \& Image Trimming\\
{\bf Stray Light Removing}\\
~~~~2. Image grouping according to ``AZIMUTH'', ``ELE'' and time period\\
~~~~3. Combine images of a group adopting ``median'' algorithm to make stray light pattern template\\
~~~~4. Each image subtract its stray light template\\
{\bf Calibration}\\
~~~~5. Astrometry\\
~~~~6. Flat Fielding With Rectified Flat Image\\
7. Source Extraction with DETECT\_THRESH=2 and DETECT\_MINAREA=4\\
{\bf Profile measurement and Clipping}\\
~~~~8. Clip objects in margins of images and clip ELONGATION$>$2 objects\\
~~~~9. Measure Moffat profile FWHMs for every objects in an image\\
~~~~10. Keep objects that have 1.3$<$MFWHM$<$3.2\\
~~~~11. Determine typical MFWHM for an image group and assign to FWHM$_{\rm med}$\\
{\bf Photometry}\\
~~~~12. Aperture photometry with apertures radii in units of FWHM$_{\rm med}$\\
~~~~13. PSF photometry\\
~~~~14. Aperture Correction\\
15. Calculate center and corners' J2000 coordinates and write into FITS header\\
16. Data and Catalogues Archiving, light curve output\\
\hline
\end{tabular}
\label{Tab:PipelineDetail}
\end{table*}

LUT’s raw data are originally obtained from the Data Management Subsystem (DMS) of Ground Research and
Application System (GRAS) of CE-3, in Level 0B, binary format\ \citep{Tan2014RAA}.
After data delivering, all data are converted to FITS (Flexible Image Transport System) format.
The FITS headers record the LUT observation modes and instrument working status.
The keywords in the headers include ``TASKCODE'' -- the type of observation task,
``IMAGETYP'' -- the type of images,
``ELE'' and ``AZIMUTH'' -- the pointing coordinates of the flat mirror (for details see Table~\ref{tab:keyword}).
\begin{table*}[!hbp]
 \centering
 \caption[]{FITS header keywords recording the LUT observation modes and instrument work status.}
 \begin{tabular}[b]{|l|l|l|} \hline
  {\bf Keyword} & {\bf Value} & {\bf Comment}\\ \hline
  \multirow{3}*{TASKCODE} & ``Initial'' & Default observation task\\ \cline{2-3} & ``Pointing'' & The pointing observation task \\ \cline{2-3} & ``Astrometry'' & Astrometry calibration observation task for the telescope\\ \cline{2-3} & ``Survey'' & The survey observation task\\ \hline
  \multirow{4}*{IMAGETYP} & ``Object'' & Image of celestial objects observation\\ \cline{2-3} & ``Zero'' & Image of zero calibration\\ \cline{2-3} & ``Dark'' & Image of dark calibration\\ \cline{2-3} & ``Flat'' & Image of flat field calibration\\ \hline
  ELE & number in arcsec & Altitude angle of the flat mirror\\ \hline
  AZIMUTH & number in arcsec & Azimuth angle of the flat mirror\\ \hline
 \end{tabular}
\label{tab:keyword}
\end{table*}

\subsection{Stray Light Removing}
\label{subsec:straylight}

The LUT detection suffers from stray light problem caused by sunlight being scattered by the cabin and the telescope.
The strength and pattern of the stray light are varying, 
with ADU counts from a few thousands in common cases to a few tens of thousands, 
depending on the angular distance between the flat mirror pointing and the sun.
Fortunately, in most cases the variation of stray light evolves very little in subsequent images taken within $\sim$0.5~hour,
so a method had been developed to remove the stray light from those images, as is described below.

\begin{figure*}[!htbp]
  \centering
  \includegraphics[width=0.6256\columnwidth]{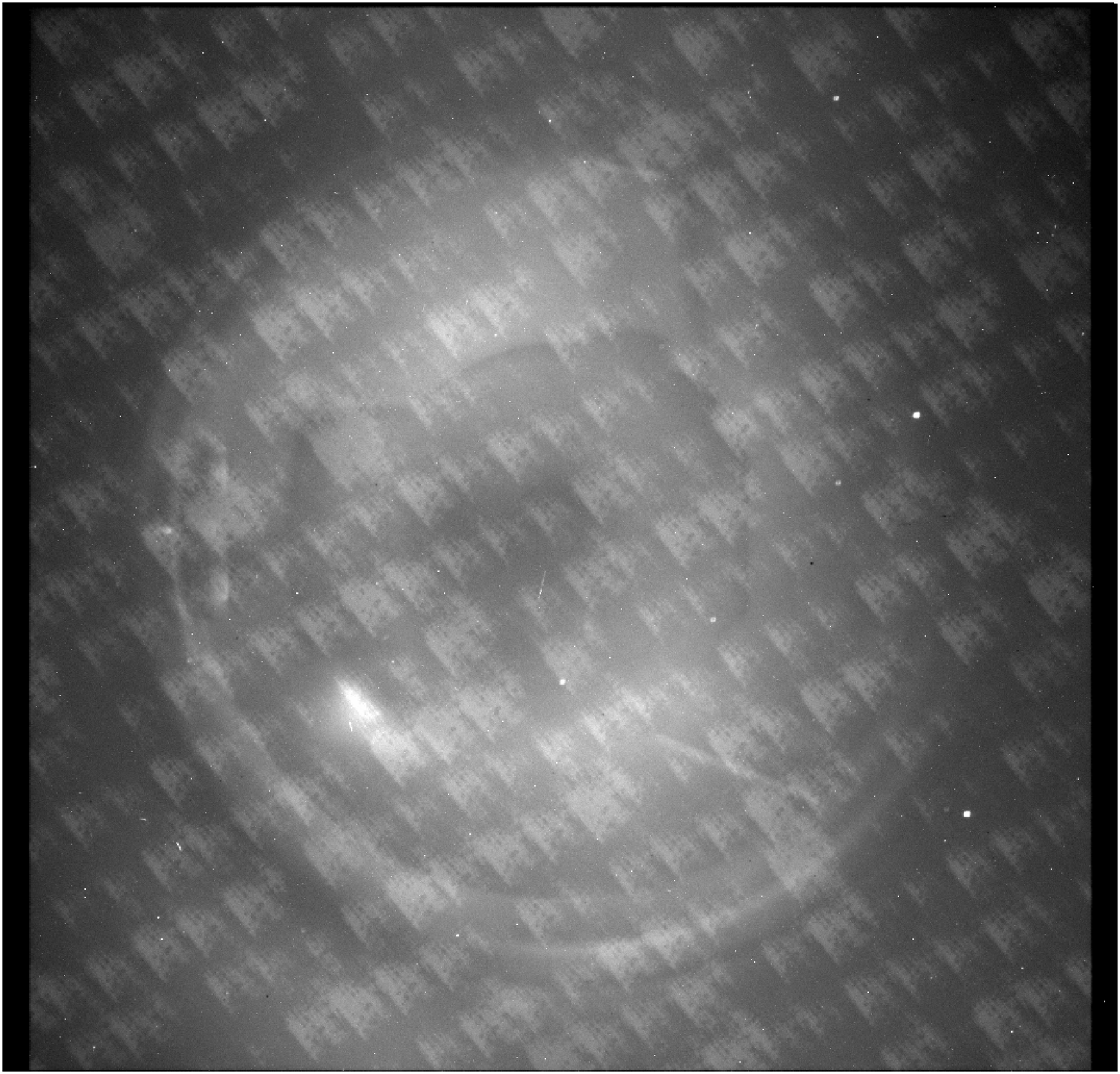}\hspace{3pt}
  \includegraphics[width=0.6\columnwidth]{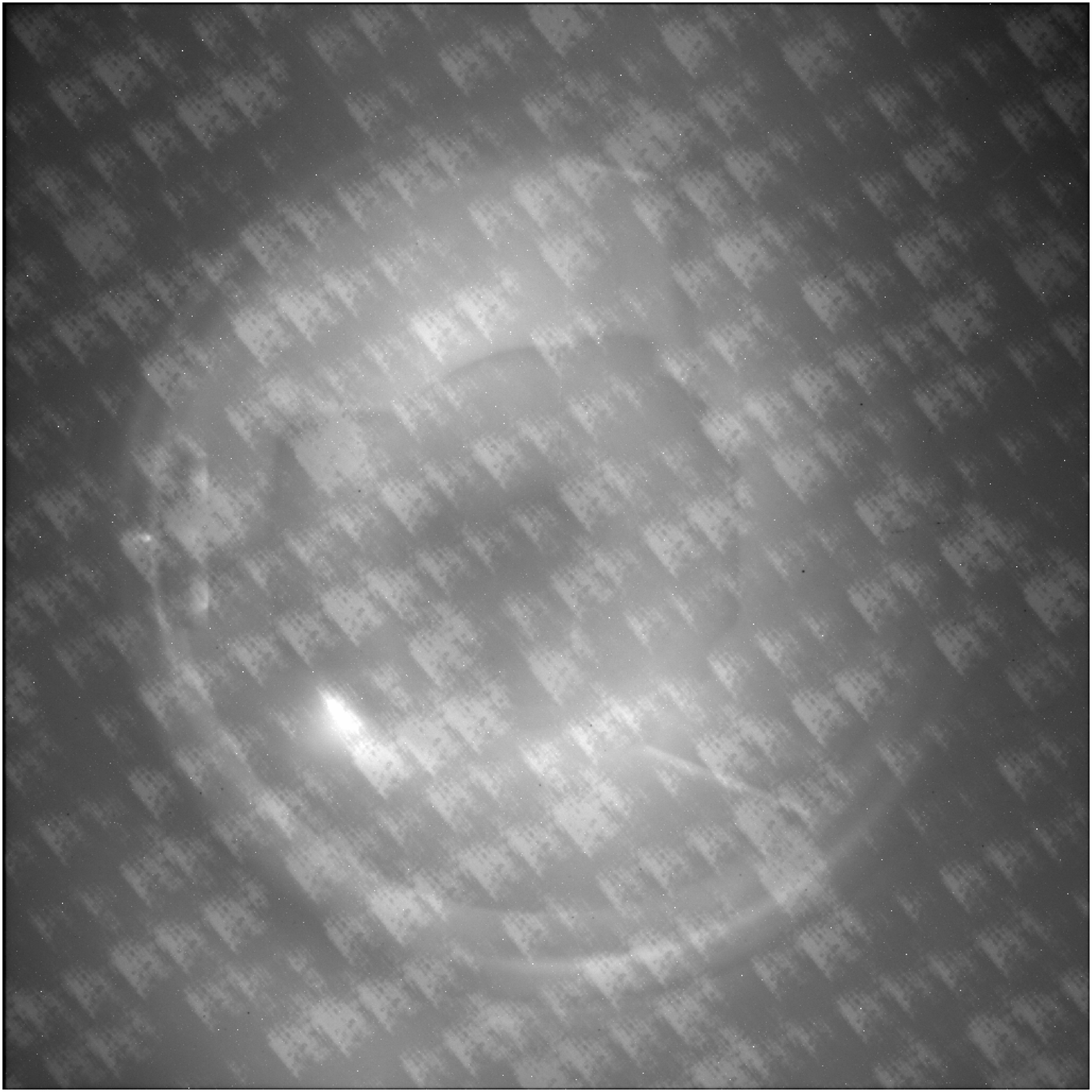}\hspace{3pt}
  \includegraphics[width=0.6\columnwidth]{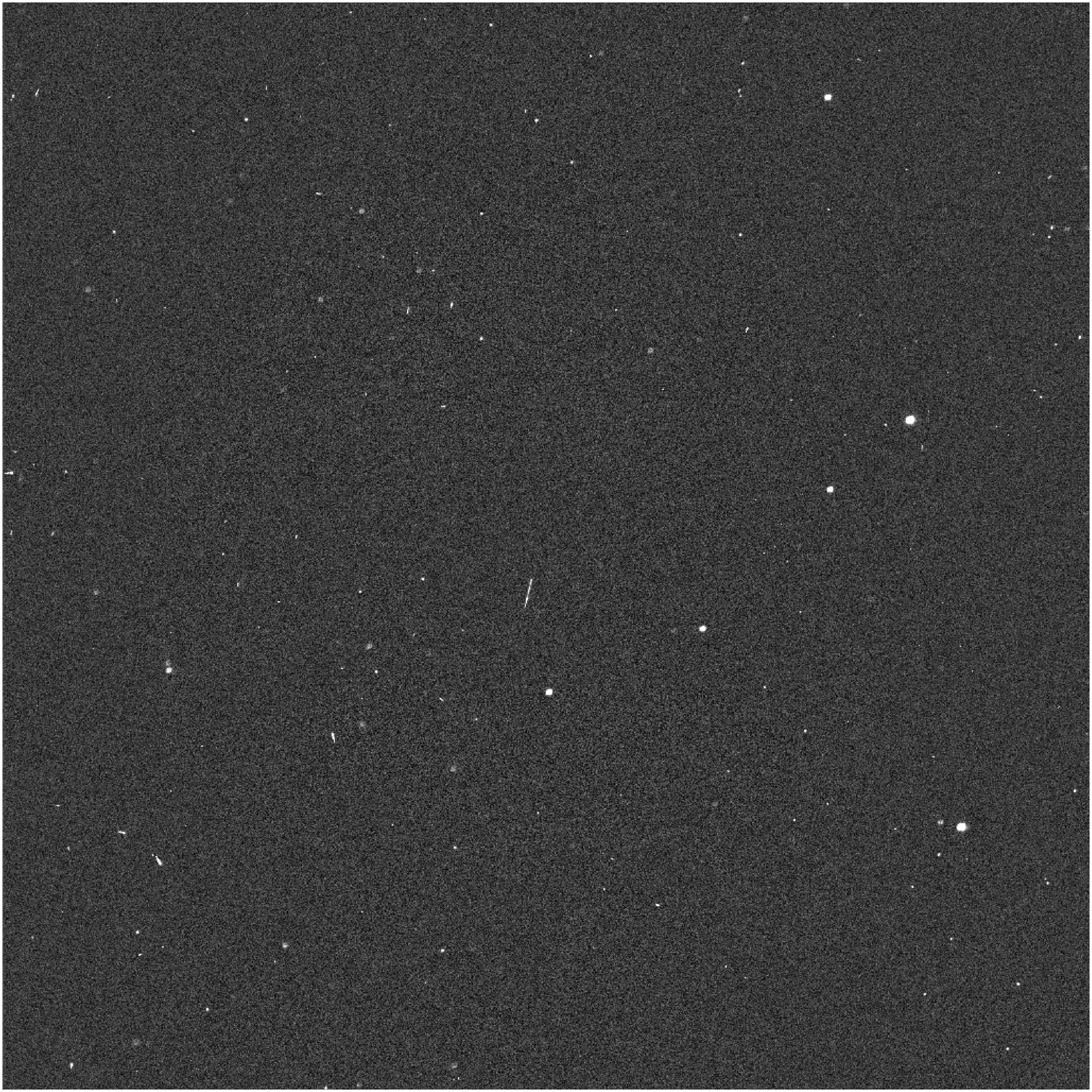}
  \caption{The effect of stray light removing. {\it Left:} An example of original LUT CCD image in size of $1072\times1027$;
           {\it middle:} The derived stray light pattern of the {\it left} image made from images of its host group in size of $1024\times1024$;
           {\it right:} Image after stray light removing in size of $1024\times1024$.}
  \label{fig:stray}
\end{figure*}

Firstly, all images are preprocessed through overscan correction and trimming to size of $1024\times1024$.
Then, images are grouped according to their head keywords ``TASKCODE'', ``ELE'', ``AZIMUTH'',
and the exposure time of each group should be within less than 1900 seconds.
For each given image, a specific stray light pattern is derived from the other images in the same group except itself through image combination using the {\sc IRAF} ``median'' algorithm. 
According to the pointing observation strategy, the stars positions on successive images 
always have slight shifts of a few pixels,
so during the ``median'' combination all celestial sources are rejected and a stray light pattern is hence left in the combined image. 
Also contained in the combined image are the underlying bias level and dark current counts. Then, each image subtracts the combined image, thereby removing the stray light pattern, bias, and dark counts.
A group commonly contains 15--30 images, so the combination can give high SNR stray light templates.
Thus, the pattern removing procedure can be considered not to induce extra noise to images.
Our stray light removing procedure also removes underlying bias and dark current in the mean time.
Figure~\ref{fig:stray} gives an example of stray light pattern and the result after its removal. 

\subsection{Astrometry}
\label{subsec:astrometry}

The astrometry is performed by cross matching the star distributions on LUT images with the 
Tycho-2 catalogue \citep{Tycho2AA}, which have been trimmed to match LUT's total available sky area.
Out of every frame 5--10 bright stars are extracted using {\sc SExtractor}
and fast photometry of the stars are performed.
The geometrical distribution and the measured LUT magnitudes of the bright stars are then cross matched to the Tycho-2 catalogue.
The star coordinates in the catalogue have been transformed to the current epoch and 
have also been corrected for precession. 
Tycho-2 optical magnitudes have to be converted to LUT NUV magnitudes for brightness matching. 
They are firstly transformed to standard Vega B, V magnitudes through 
the transformation relationships provided by the {\it Hipparcos and Tycho Catalogues} \citep{Tycho1997}.
Then, the Vega B, V magnitudes are used to calculate theoretical LUT AB magnitudes through the stellar atmosphere model from 
\citet{Kurucz2004astroph} for a series of different spectral types (details are described in a paper of Han\ et\ al. in preparation). 
For each matched star, in addition to the cross matching radius constraint,
the difference between the measured and calculated LUT magnitude is used as further constraint 
and is required to be less than $\sim$2\,mag.
If the cross matching successes, the world coordinate system (WCS) (J2000) as a result is written into the FITS header.
If the matching fails, the pipeline gives up the current frame and jumps to the next one.
If more than 20\% of total frames fail in the matching, the automatic data processing pipeline working with WCS terminates and 
an alternative pipeline working without WCS will be carried out.
The accuracy of astrometry is typically about 1''.

\subsection{Flat Fielding}
\label{subsec:flatcorr}

\begin{figure*}[!htbp]
  \centering
  \includegraphics[width=\columnwidth]{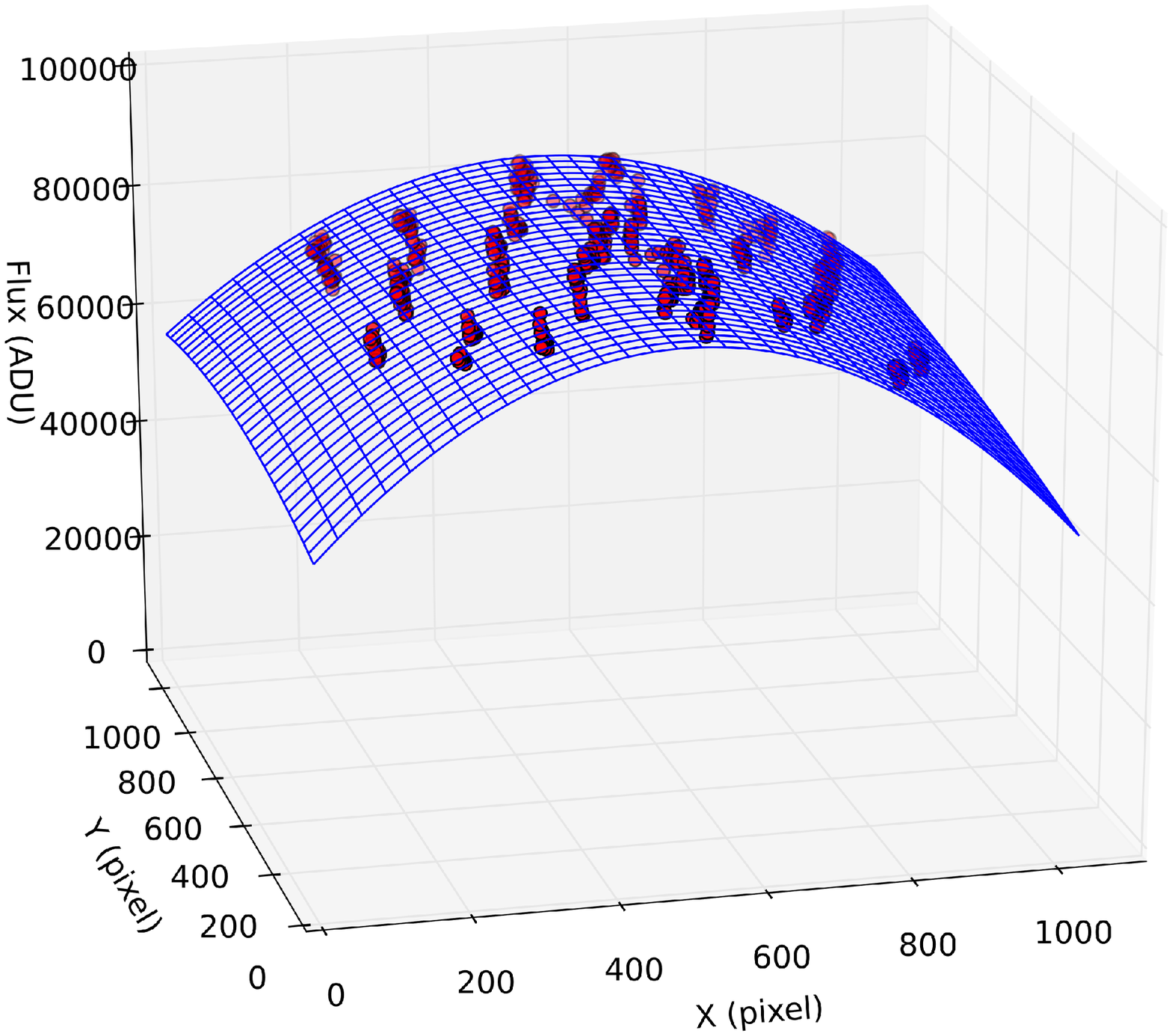}\\
  \includegraphics[width=0.6\columnwidth]{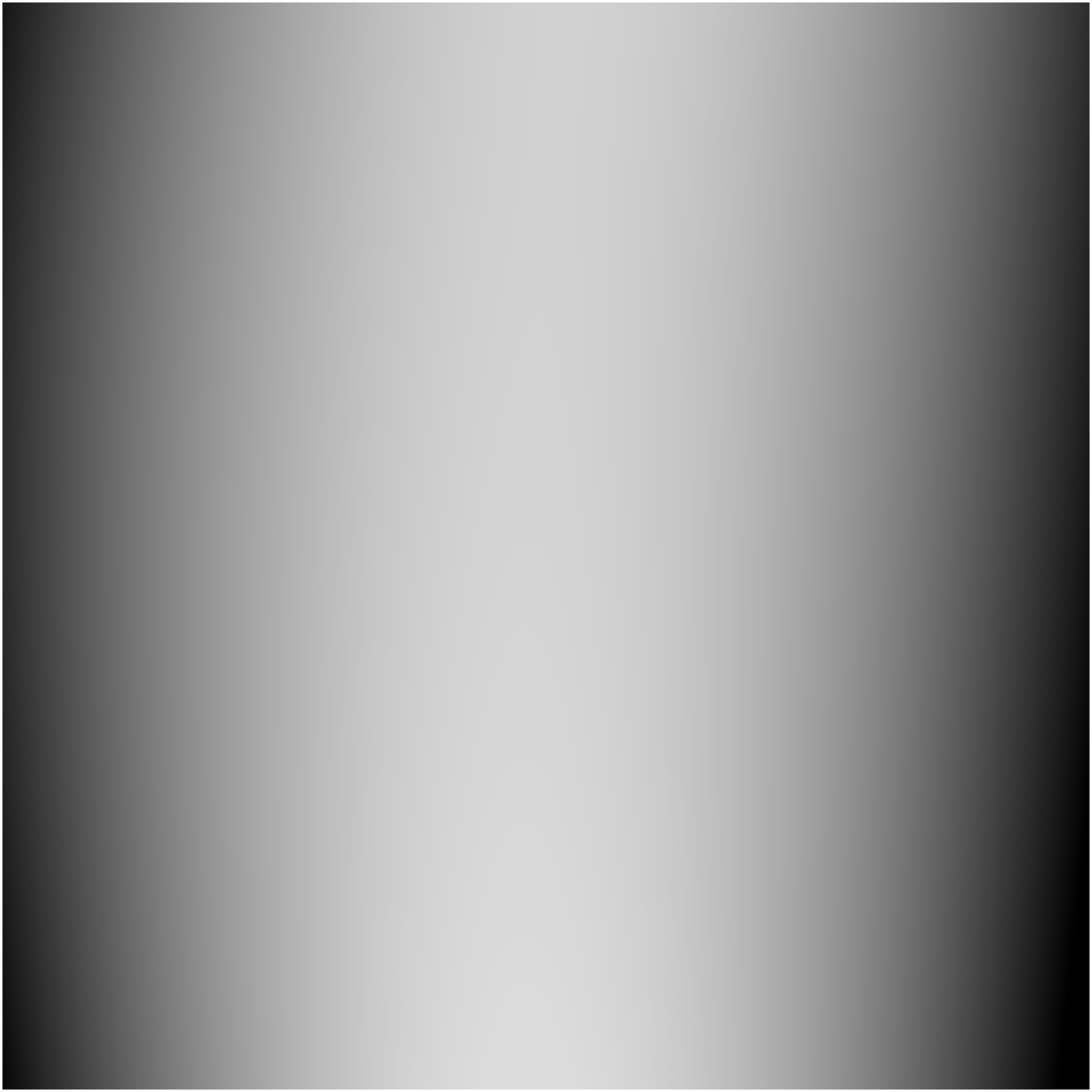}\hspace{5pt}
  \includegraphics[width=0.6\columnwidth]{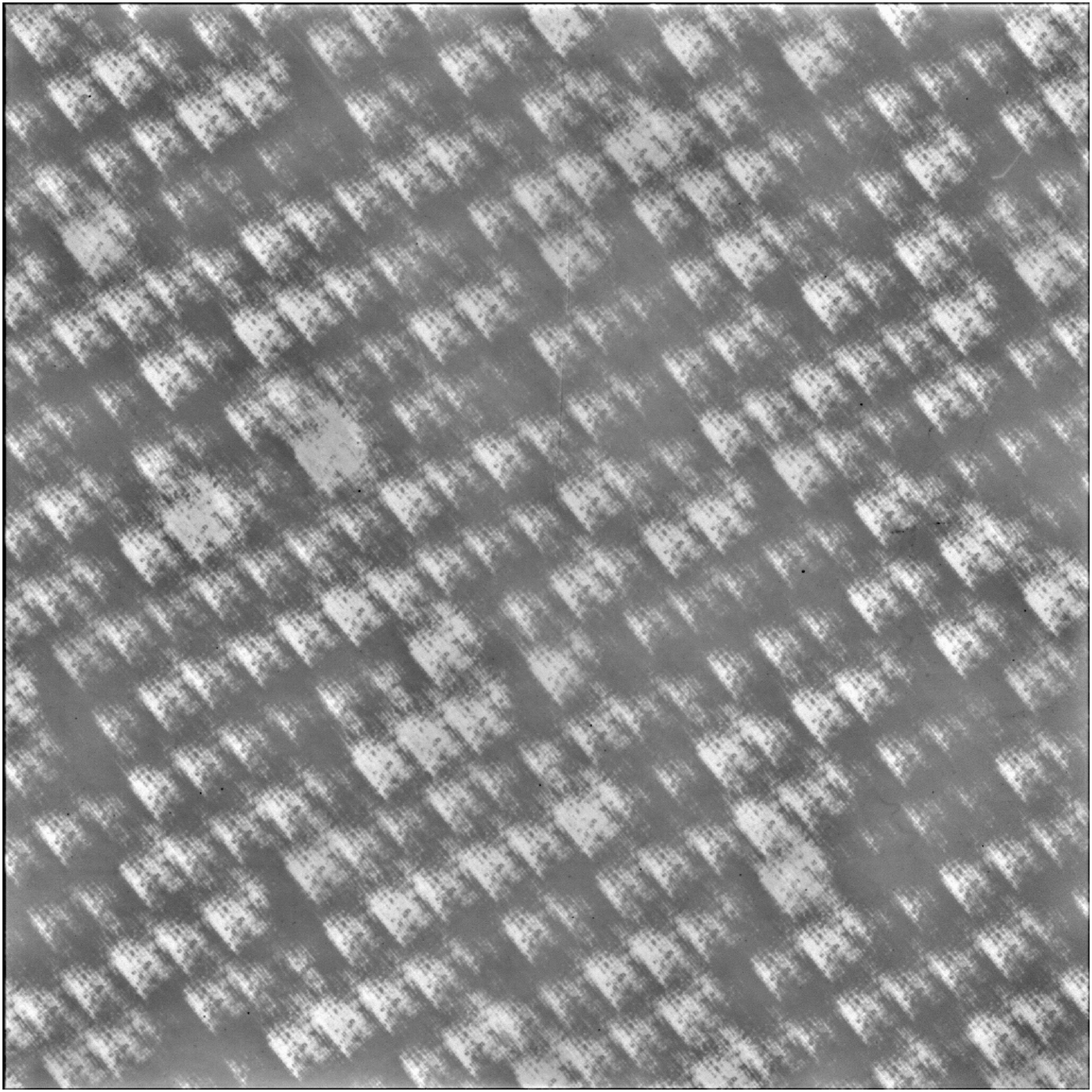}\hspace{5pt}
  \includegraphics[width=0.6\columnwidth]{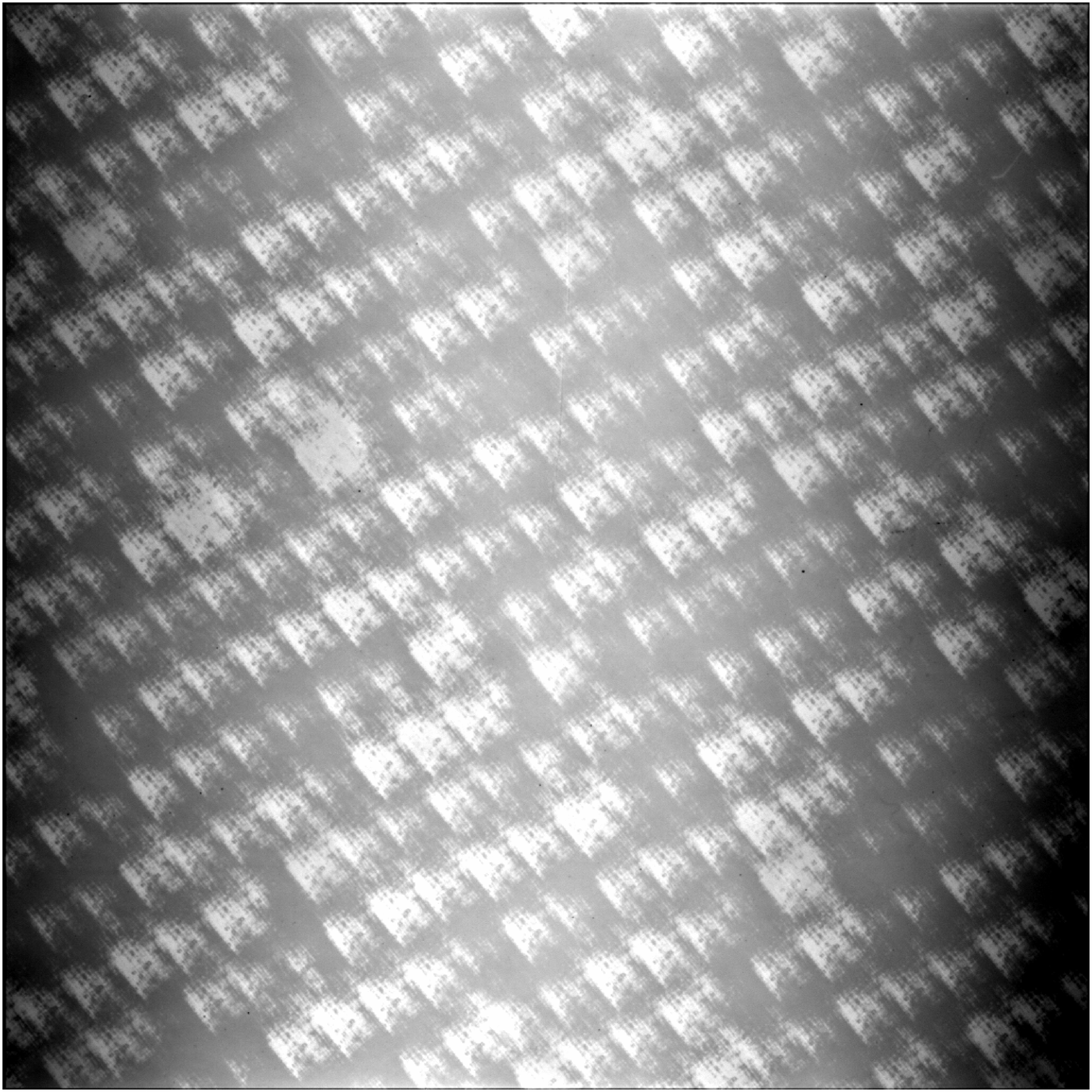}
  \caption{Flat field creation. {\it Top:} two-dimensional, second-order polynomial fitting to superflat;
          {\it bottom left:} an image of the two-dimensional, second-order polynomial fitting to superflat (a);
          {\it bottom middle:} the processed internal flat field image only retaining pixel-to-pixel response nonuniformity (b);
          {\it bottom right:} final flat field image for calibration = (a)$\times$(b).}
  \label{fig:superflat}
\end{figure*}

The flat fielding procedure makes use of both the internal flat field from internal LEDs and superflat
from dithering observations (see Sect.~\ref{sec:Obser}).
The internal flat field has been processed to retain only the CCD pixel-to-pixel response nonuniformity, 
filtering out the low-frequence structure, which produces illumination corrected and normalized LED flat field.
The dithering observation of a single standard star produces a positional sampling grid of LUT's large-scale response nonuniformity.
These nonuniformity structures are recorded in both the background and the stars in each image.
After stray light removing, the information of large-scale response in background is removed, but is left in stars' fluxes. 
Since each standard star is used as an ``invariable'' light source, its fluxes at different positions in the 
FOV should exhibit the structure of large-scale response nonuniformity.
The flux counts at these grid positions are measured through aperture photometry, 
before which, flat fielding using the internal flat field image is performed.
Then, the fluxes of the grid is fitted by a two-dimensional, second-order polynomial function and the superflat is created
(see Fig.~\ref{fig:superflat} {\it top}).
The final flat field that is used in the pipeline is the product of image multiplication of the superflat and the illumination corrected and normalized LED flat field 
(see Fig.~\ref{fig:superflat} {\it bottom}).

\subsection{Source Extraction, Brightness Profile Measurement and Cosmic Ray Rejection}
\label{subsec:sextractor}

Sources in each frame are extracted using {\sc SExtractor} by certain criteria 
to conveniently obtain their CCD X-Y coordinates.
The key extraction criteria are summarized in~Table~\ref{tab:SEconfig}.
The selected parameters for {\sc SExtractor} output are X\_IMAGE (object's barycenter position along X-axis), 
Y\_IMAGE, ELONGATION (shape parameter, major-axis/minor-axis), BACKGROUND, {\it etc}.

\begin{figure}[!htbp]
 \centering
 \includegraphics[width=\columnwidth]{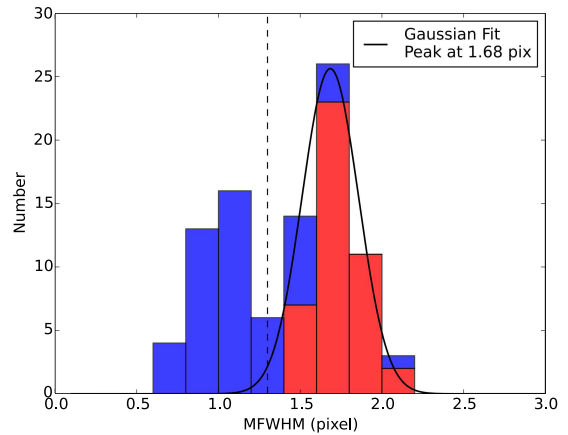}
 \caption{An example of histogram statistics of MFWHMs of celestial objects ({\it red}) and cosmic rays ({\it blue}) for a single image.
          A clipping line is set at MFWHM=1.3 to remove most cosmic rays from the extracted catalogue.
          A 1D gaussian fit is performed ({\it black} line) to the histogramic distribution profile after 1.3$\leq$MFWHM$\leq$3.2 filtering.}
 \label{fig:clip}
\end{figure}

In order to determine the aperture size for photometry (see Sect.~\ref{subsec:phot}),
brightness profile measurement and cosmic ray rejection are performed for each image.
First of all, sources in edge regions are cleaned from the extracted catalogue because they may be 
signals arose just by instruments.
Then, FWHMs of sources profiles are measured by the ``psfmeasure'' task in {\sc IRAF}.noao.obsutil package. 
{\it X-} and {\it Y-}axis positions in the output file from {\sc SExtractor} are used as inputs of the task,
and Moffat profile widths (denoted as MFWHMs) are required as output.
In LUT images, cosmic ray events are comparable to celestial sources in number.
The event rate of cosmic rays is estimated to be $\sim$4 cosmic rays per second in the FOV,
corresponding to $\sim$2 cosmic rays per second per square degree.
We adopt the cosmic ray rejection criterion as MFWHM$<$1.3.
The criterion is determined through sources identification in some typical cases,
in which cosmic rays are identified by correlation of successive images.
Figure~\ref{fig:clip} shows the MFWHM distributions of the identified stars ({\it red}) and cosmic rays ({\it blue}) in a typical case.
Although a few cosmic rays are blended with stars in terms of their profile,
the distributions of the two populations are obviously seperated, which enable us to reject most cosmic rays
in terms of MFWHM$\sim$1.3.
Objects with MFWHM$>$3.2 are deemed as extended sources or cosmic ray clumps, so they are also clipped off from the extracted catalogue.
Residual cosmic rays are further rejected with the shape criterion of ELONGATION$>$2, 
if the target of pointing observation is not a binary or blended stars.
The clipping criteria are summarized in Table~\ref{tab:SEconfig}.

\begin{table*}[!htbp]
\centering
\caption[]{Source extraction criteria for {\sc SExtractor} and clipping}
\begin{tabular}{|l|c|}
\hline
\multicolumn{2}{|c|}
{\bf Source extraction criteria for {\sc\bf SExtractor}}\\
\hline
{\bf Criteria} & {\bf {\sc\bf SExtractor} parameters set up}\\
\hline
Detection threshold relative to background RMS & DETECT\_THRESH = 2\\
Minimum number of connected pixels above threshold & DETECT\_MINAREA = 4\\
Background mesh size & BACK\_SIZE = 64\\
\hline
\multicolumn{2}{|c|}
{\bf Clipping Criteria For Output Parameters of {\sc\bf SExtractor} And {\sc\bf IRAF}.psfmeasure}\\
\hline
{\bf Criteria} & {\bf Parameters set up}\\
\hline
Clip edge region of CCD columns & 20 $<$ X\_IMAGE $<$ 1020\\
Clip edge region of CCD rows & 5 $<$ Y\_IMAGE $<$ 1020\\
Clip stretched objects & ELONGATION $<$ 2\\
Clip cosmic rays and extended sources & 1.3 $<$ MFWHM $<$ 3.2\\
\hline
\end{tabular}
\label{tab:SEconfig}
\end{table*}

\subsection{Aperture and PSF Photometry}
\label{subsec:phot}

Both aperture and PSF photometries are performed for all the extracted objects after the cosmic ray rejection. 
The AB magnitude system of \citet{Oke1983} is adopted following 
Sloan Digital Sky Survey (SDSS, \citealt{Fukugita1996AJ}), Galaxy Evolution Explorer (GALEX, \citealt{Morrissey2007ApJS}), {\it etc}.
LUT magnitude is defined as
\begin{equation}
 m_{\rm LUT} = m_{\rm 0,LUT} - 2.5 \log f_{\rm LUT}
\end{equation}
where $m_{\rm 0,LUT}$ is the zero point magnitude of LUT,
$f_{\rm LUT}$ is the flux density in ergs\,s$^{\rm -1}$\,cm$^{\rm -2}$\,Hz$^{\rm -1}$).
Zero point magnitude of photometry is calibrated for LUT system as
\begin{equation}
m_{\rm 0,LUT} = 17.52\pm0.05.
\end{equation}
Details of LUT photometry calibration are given by \citet{Wangjing2015cali}. 
The corresponding signal-to-noise ratios (SNRs) are calculated as:
\begin{equation}
 {\rm SNR}_{r_{\rm i}} = \dfrac{F_{r_{\rm i}}}{\sqrt{\dfrac{F_{r_{\rm i}}}{G} + A \times \sigma^2 + \dfrac{A^2\times\sigma^2}{N_{\rm sky}}}}
\end{equation} 
({\sc IRAF}.apphot.phot HELP document) where $F_{r_{\rm i}}$ is the total number of counts excluding background in aperture $r_{\rm i}$, 
G is the gain of CCD (electrons per ADU), 
A is the area in aperture $r_{\rm i}$ in square pixels, 
$\sigma$ is the standard deviation of the background which mainly includes readout noise, stray light noise, bias and dark counts noise, 
local flat fielding noise, etc., 
$N_{\rm sky}$ is the pixels number of background. 

The aperture photometry is performed with a series of aperture sizes, whose radii are
1$\times$, 1.5$\times$, 2$\times$, 2.5$\times$, 3$\times$, and 4$\times {\rm FWHM_{med}}$,
where ${\rm FWHM_{med}}$ is denoted as the stars' typical FWHM.
The ${\rm FWHM_{med}}$ is calculated for each image group which has formed in the stray light removing procedure.
The calculation method of ${\rm FWHM_{med}}$ is as follows.
Firstly, for each image, MFWHMs that satisfy 1.3$\leq$MFWHM$\leq$3.2 are plotted in a histogram with bar width of 0.2~pixel.
Secondly, a typical FWHM for each image is obtained by fitting the MFWHM distribution 
with a one-dimensional Gaussian function and deriving the Gaussian peak value (denoted as MFWHM$_{\rm peak}$).
Instead of a PSF measurement based on the brightest star,
this statistical method is adopted because there are some cosmic ray events whose 
profiles perfectly mimic that of a bright star, which can not be removed through the rejection 
criteria described above. 
Thirdly, the median value of MFWHM$_{\rm peak}$ of a group of images 
(15--30 in number varies for different pointing observation tasks) 
is derived and denoted as ${\rm FWHM_{med}}$, and is used as the unit of aperture size.
The average value (``median'' algorithm) is adopted here to deal with possible star brightness profile variation,
although such variation is negligible (within $\sim$0.04~pixel through an observation task) for most cases.
The variation may only be significant (in a factor of $\sim$2) in the first observational day of each month,
if the telescope has not reached its designed thermal equilibrium state.
Background annulus for aperture photometry is set to be 6$\times$ and 8$\times {\rm FWHM_{med}}$ as inner and outer radii, respectively.
An input file of sources positions for each image is used as {\sc IRAF} image cursor,
so aperture photometry runs automatically.
The photometric error is typically $\sim$0.01~mag for LUT band 10~mag stars (30s exposure),
which comes from image background noises.

\begin{figure*}[!htbp]
  \centering
  \includegraphics[width=0.9\columnwidth]{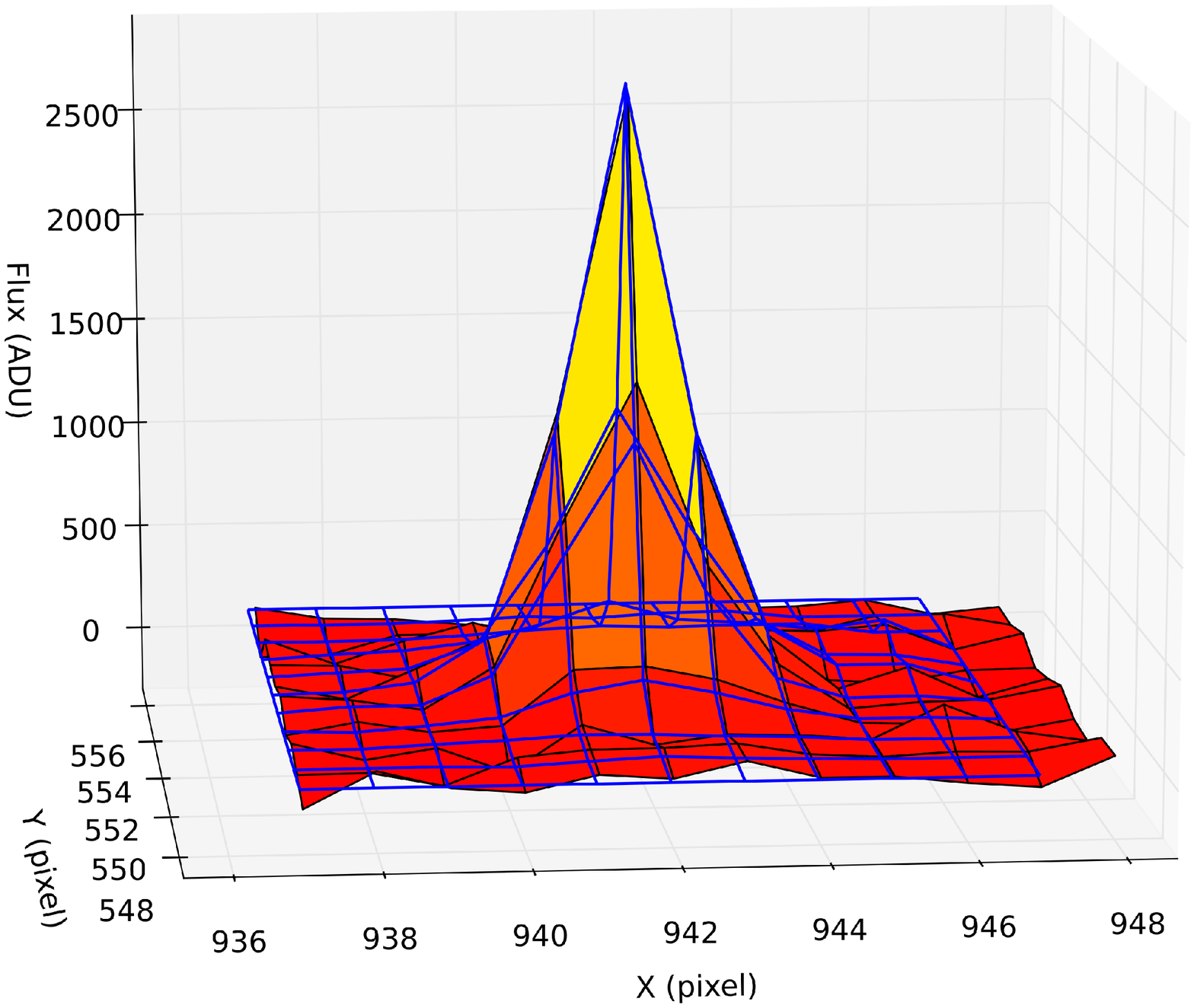}
  \includegraphics[width=0.9\columnwidth]{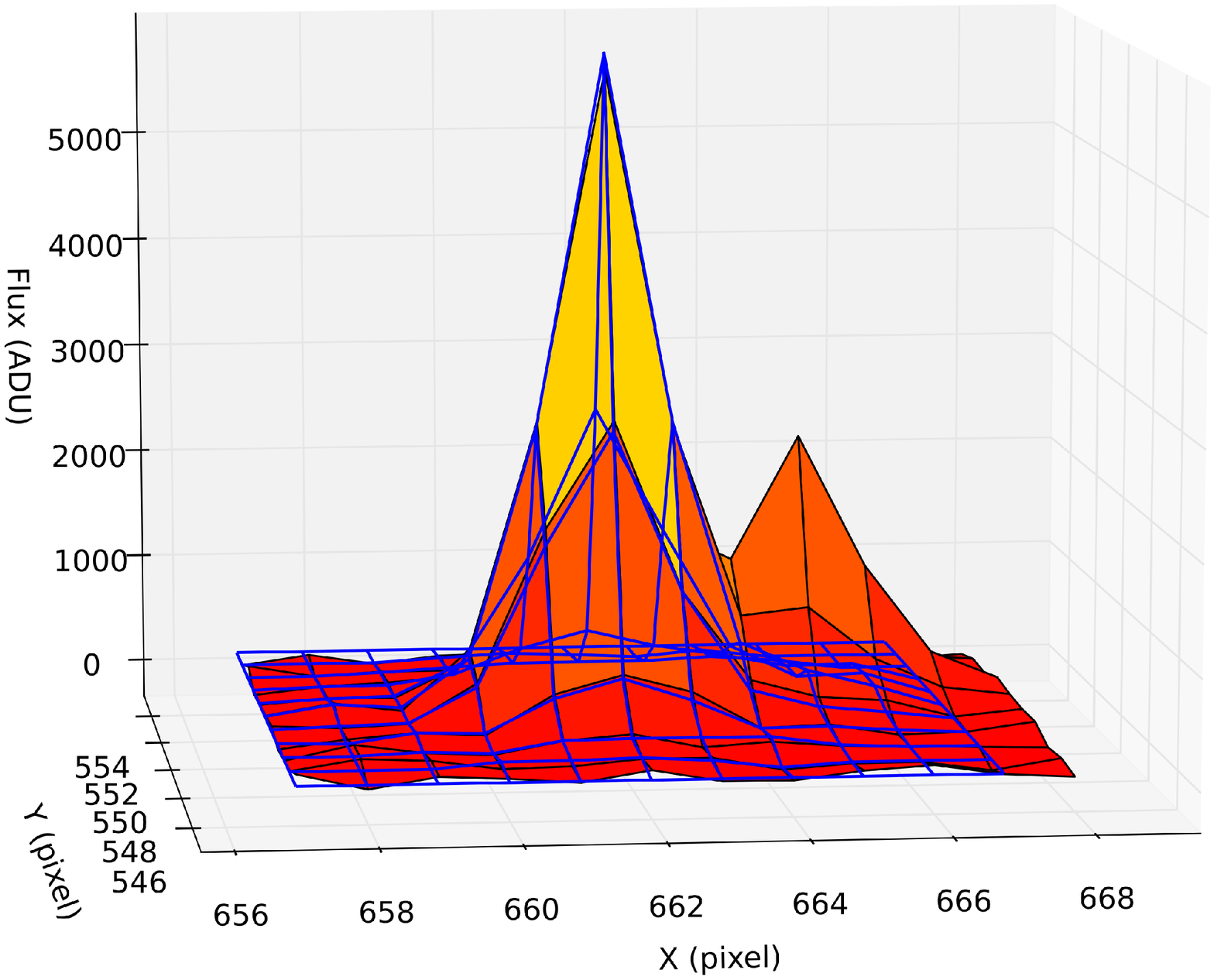}
  \caption{PSF fitting for single and double star. The {\it hot} colored surfaces illustrate source fluxes, 
           and the {\it blue wired} frames illustrate PSF models to the sources.
           {\it Left:} PSF fitting with Moffat function for a star with SNR$\sim$31;
           {\it right:} PSF fitting with Moffat function for a star with SNR$\sim$43 in a binary star BW\,Dra.}
  \label{fig:psffit}
\end{figure*}

Before PSF photometry, a further clipping based on FWHM criterion is carried out to select candidate PSF stars.
We firstly clipped the extracted sources with FWHM$<$1.4 and FWHM$>$2.5\,pixels. After the clipping, 
the brightness profiles of the 10 brightest objects are fitted through $\chi^2$ minimization. 
A test work of comparing various profile functions, such as Gaussian, Lorentzian and Pennian functions, 
has been carried out and indicates that an elliptical Moffat function with a fixed $\beta$ parameter of 1.5 provides best fits more frequently.
After the PSF model establishment, PSF photometry is carried out within circle radii of 3$\times {\rm FWHM_{med}}$.
The inner radius of the annulus used to determine background level is 5$\times {\rm FWHM_{med}}$, 
and the annulus width is 2$\times {\rm FWHM_{med}}$. 
The effects of PSF fitting for single and double star are shown in Figure~\ref{fig:psffit}.

\section{Aperture Correction}
\label{sec:Apertcor}

\begin{figure*}[!htbp]
  \centering
  \includegraphics[width=0.9\columnwidth]{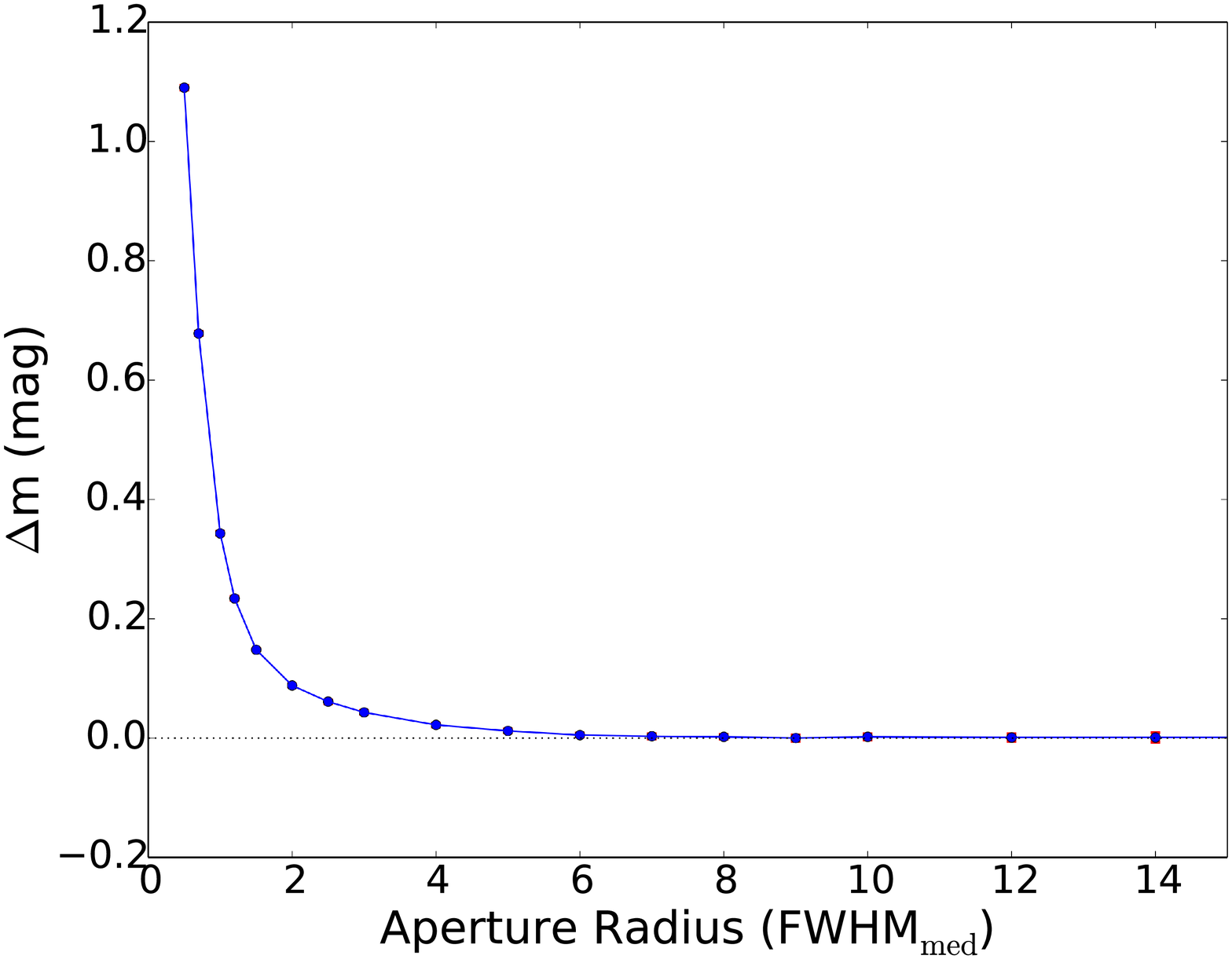}
  \includegraphics[width=0.9\columnwidth]{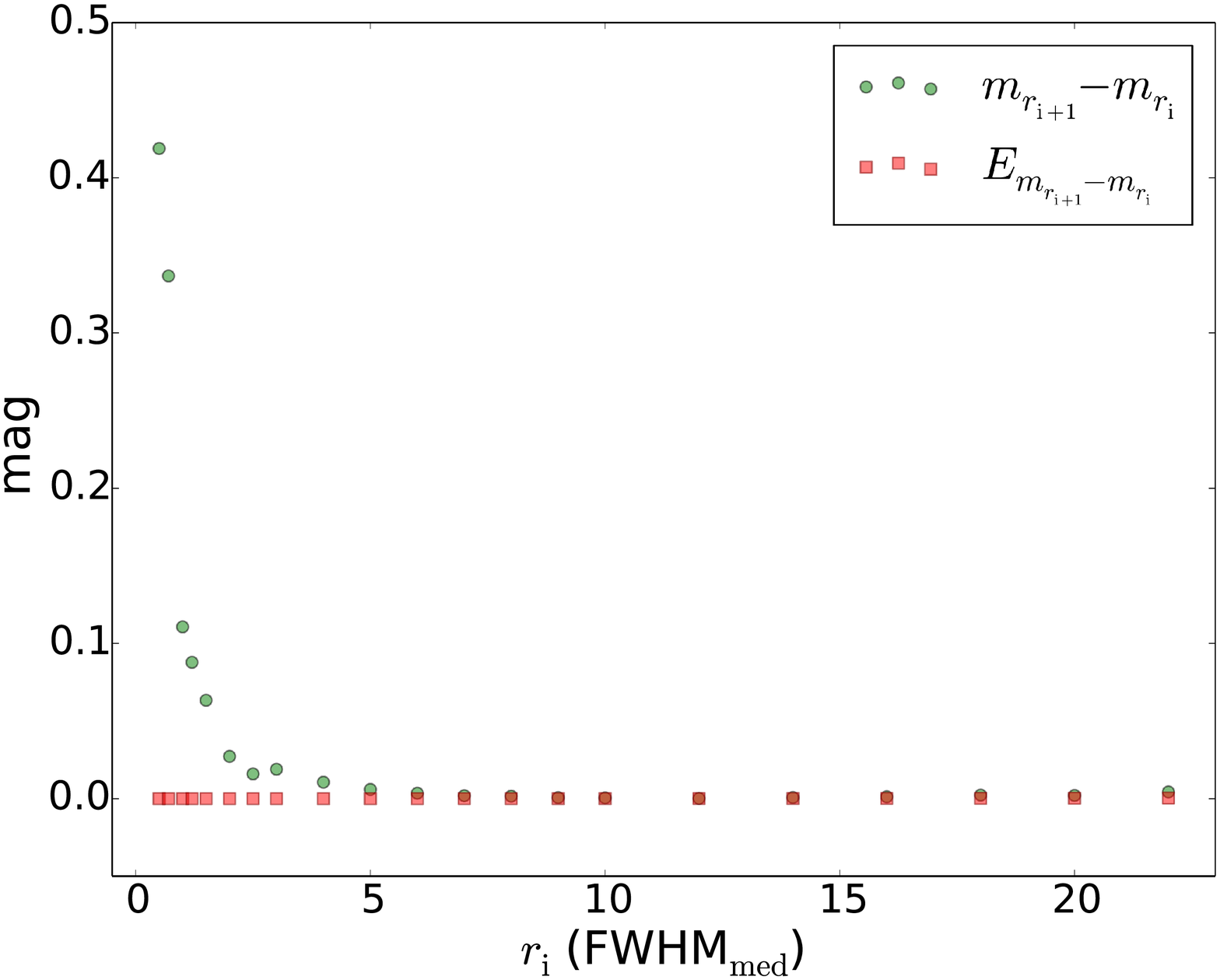}
  \caption{{\it Left:} LUT point source photometric curve of growth. Magnitude of 14$\times{\rm FWHM_{med}}$ aperture is adopted as 
             the reference line, and aperture axis maximum range for clear show;
           {\it right:} $|m_{r_{\rm i+1},{\rm med}}-m_{r_{\rm i},{\rm med}}|$ ({\it green circles}) and the corresponding errors ({\it red squares}).}
  \label{fig:growth}
\end{figure*}

The correction of aperture effect is applied to the magnitudes measured by the aperture and PSF photometry to
obtain the magnitude measured in an ``infinite'' size aperture, which is considered to embrace the total flux of a source.
The aperture correction is performed by
\begin{equation}
 m_{r_{\rm i},{\rm cor}} = m_{r_{\rm i}} + \Delta m_{r_{\rm i}}
\end{equation}
where $\Delta m_{r_{\rm i}}$ is the aperture correction factor for aperture $r_{\rm i}$,
$m_{r_{\rm i}}$ and $m_{r_{\rm i},{\rm cor}}$ are magnitudes before and after aperture correction, respectively.
The values of the correction factors are determined through the ``curve of growth'' method. 
The curve of growth is derived by performing aperture photometry for standard star HD\,185395 with a dense sampled aperture radii series.
The 22 sampled aperture radii range from 0.5$\times {\rm FWHM_{med}}$ to 25$\times {\rm FWHM_{med}}$. 
The standard star has 43 frames of exposure, and 43 curves of growth are obtained through aperture photometry.
Mean values of magnitudes at each aperture radii are calculated to derive the mean curve of growth.
Figure~\ref{fig:growth} ({\it left}) shows the mean curve of growth in terms of the magnitude offsets relative to magnitude measured within radius of 14$\times{\rm FWHM_{med}}$. 
The solid line in Figure~\ref{fig:growth} ({\it left}) is the modeling approach to the mean curve of growth, 
which can be expressed by the following function:
\[ \Delta m(r) = \left\{ \begin{lgathered}
            1-2.5\log(1-e^{-\frac{r}{1.93-2.29r+1.4r^2-0.27r^3}}),\\
             \phantom{MMMMMMMMMMMMM}(r \leqq 2)\\
            1.22-0.1r+0.016r^2-0.00086r^3,\\
             \phantom{MMMMMMMMMMMMM}(r > 2)
\end{lgathered} \right. \]
where $r$ is the desired aperture radius to which the photometry is corrected in units of FWHM$_{\rm med}$.

Aperture correction factors are calculated as mean differences between magnitudes in every apertures
and the magnitude in the aperture from which magnitude differences converge to zero and also converge to their errors.
The criteria of converge are
\begin{eqnarray}
|m_{r_{\rm i+1}}-m_{r_{\rm i}}| & \simeq & 0 \\
|m_{r_{\rm i+1}}-m_{r_{\rm i}}| & \lesssim & E_{m_{r_{\rm i+1}}-m_{r_{\rm i}}}
\end{eqnarray}
where $m_{r_{\rm i}}$ is the magnitude measured in aperture $r_{\rm i}$ of the mean curve of growth, \\
$E_{m_{r_{\rm i+1}}-m_{r_{\rm i}}}$
is the error of $m_{r_{\rm i+1}}-m_{r_{\rm i}}$.
Figure~\ref{fig:growth} ({\it right}) shows the magnitude differences and their errors versus aperture radii.
From 8$\times{\rm FWHM_{med}}$ on and up to 16$\times{\rm FWHM_{med}}$ aperture radius, the magnitude differences converge.
A mean value of magnitudes in 10$\times$, 12$\times$, and 14$\times{\rm FWHM_{med}}$ aperture radii is calculated 
and used as the magnitude of total flux.
Magnitude offsets between 1$\times$, 1.5$\times$, 2$\times$, 2.5$\times$, 3$\times$, and 4$\times{\rm FWHM_{med}}$ 
aperture photometry magnitudes and the total-flux-magnitude 
are used as aperture correction factors, whose values are listed in Table~\ref{tab:apercorr}.

\begin{table}[!htbp]
\centering
\caption[]{Aperture correct factors (in magnitude) and their errors.}
\label{tab:apercorr}
\setlength{\tabcolsep}{1pt}
 \begin{tabular}{cc}
  \hline\noalign{\smallskip}
  $r_{\rm i}$ & $\Delta m_{r_{\rm i}}$\\
  ({\footnotesize ${\rm FWHM_{med}}$}) &\\
  \hline\noalign{\smallskip}
  1.0 & -0.348 $\pm$ 7.3E-4\\
  1.5 & -0.150 $\pm$ 6.5E-4\\
  2.0 & -0.086 $\pm$ 6.5E-4\\
  2.5 & -0.059 $\pm$ 7.2E-4\\
  3.0 & -0.043 $\pm$ 7.3E-4\\
  4.0 & -0.024 $\pm$ 7.3E-4\\
  \noalign{\smallskip}\hline
\end{tabular}
\end{table}

\section{Pipeline Performance}
\label{sec:Discussion}

\begin{figure}[!htbp]
 \centering
 \includegraphics[width=\columnwidth]{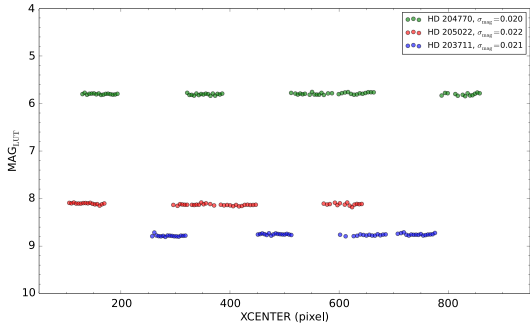}
 \caption{Photometry test for integrated effect of stray light removing, flat fielding and aperture photometry.}
 \label{fig:flattest}
\end{figure}

We have carried out some test work to assess the accuracy of the data processing pipeline.
Since the flat field images are made from the processed internal flat field images retaining pixel-to-pixel nonuniformity,
and the superflat reflecting large-scale nonuniformity,
they would not contain medium-scale (tens to one hundred pixels) structures,
which may bring flat fielding related errors to photometry.
The influence of medium-scale structures can be tested by high frequency positional sampling observations 
whose targets go across the image frame.
At each position, aperture photometry is performed and magnitude is obtained to find out the medium-scale structures effect.
Figure~\ref{fig:flattest} shows the result of such test carried out in June, observing stars HD\,204770, 
HD\,205022 and HD\,203711, where XCENTER means X-axis positions of the targets, MAG$_{\rm LUT}$ means LUT magnitudes 
measured in 3$\times {\rm FWHM_{med}}$ aperture radius.
Standard deviations of magnitudes are 0.020\,mag for HD\,204770, 0.022\,mag for HD\,205022 and 0.021\,mag for HD\,203711.
Such dispersion is mostly caused by medium-scale nonuniformity, 
which is estimated to be $\sim$0.02\,mag after
deducting errors of image background noises ($\sim$0.01\,mag for a 10\,mag star).
The error increases to $\sim$0.2\,mag for stars of 13.5\,mag with 30\,s exposure,
and this corresponds to the 5$\sigma$ detection limit of LUT.

\section{Summary}

We describe the data processing pipeline reducing the pointing observation data of LUT.
The pipeline performs stray light removing, astrometry, flat fielding employing superflat technique,
source extraction, source profile measurement, cosmic ray rejection, 
aperture and PSF photometry, aperture correction, catalogue archiving, and outputs light curves.
The pipeline has been intensively tested and works smoothly with observation data.
The photometric accuracy is typically $\sim$0.02\,mag for LUT 10\,mag stars (30\,s exposure),
with errors from all that of background noise, residuals of stray light removing, and flat fielding.
The accuracy degrades to be $\sim$0.2\,mag for stars of 13.5\,mag which is the 5$\sigma$ detection limit of LUT.

\acknowledgments

This project is supported by the Key Research Program of Chinese Academy of Science (KGED-EW-603),
the National Basic Research Program of China (973-program, Grant No. 2014CB845800),
and the National Natural Science Foundation of China (Grant Nos. 11203033, 11473036, U1231115, U1431108). 
The authors thank Shun-Fang Liu for the help of data reduction, and Yang Xu for the help of building the LUT database.
This project made use of {\sc SExtractor}, a powerful program for astronomical data analysis \citep*{Bertin1996A&AS}.
{\sc Astropy}, a community-developed core {\sc Python} package for Astronomy ({\sc Astropy} Collaboration, 2013).
{\sc Matplotlib}, a 2D graphics package for {\sc Python} \citep{Hunter:2007}, is used in this work for figure making. 
{\sc PyRAF} and {\sc PyFITS} are products of the Space Telescope Science Institute, which is operated by AURA for NASA.
This project make a lot of use of the tabular data analysis and visualization software {\sc TOPCAT} \citep{Topcat2005}
to do archiving works, which has not been described in the text.

\bibliographystyle{spr-mp-nameyear-cnd}
\bibliography{xmmeng_bib}

\begin{thebibliography}{13}
\ifx \bisbn   \undefined \def \bisbn  #1{ISBN #1}\fi
\ifx \binits  \undefined \def \binits#1{#1} \fi
\ifx \bauthor  \undefined \def \bauthor#1{#1} \fi
\ifx \batitle  \undefined \def \batitle#1{#1} \fi
\ifx \bjtitle  \undefined \def \bjtitle#1{#1}\fi
\ifx \bvolume  \undefined \def \bvolume#1{\textbf{#1}}\fi
\ifx \byear  \undefined \def \byear#1{#1} \fi
\ifx \bissue  \undefined \def \bissue#1{#1} \fi
\ifx \bfpage  \undefined \def \bfpage#1{#1} \fi
\ifx \blpage  \undefined \def \blpage #1{#1} \fi
\ifx \burl  \undefined \def \burl#1{\textsf{#1}} \fi
\ifx \doiurl  \undefined \def \doiurl#1{\textsf{#1}} \fi
\ifx \betal  \undefined \def \betal{\textit{et al.}} \fi
\ifx \binstitute  \undefined \def \binstitute#1{#1} \fi
\ifx \binstitutionaled  \undefined \def \binstitutionaled#1{#1} \fi
\ifx \bctitle  \undefined \def \bctitle#1{#1} \fi
\ifx \beditor  \undefined \def \beditor#1{#1} \fi
\ifx \bpublisher  \undefined \def \bpublisher#1{#1} \fi
\ifx \bbtitle  \undefined \def \bbtitle#1{#1} \fi
\ifx \bedition  \undefined \def \bedition#1{#1} \fi
\ifx \bseriesno  \undefined \def \bseriesno#1{#1} \fi
\ifx \blocation  \undefined \def \blocation#1{#1} \fi
\ifx \bsertitle  \undefined \def \bsertitle#1{#1} \fi
\ifx \bsnm \undefined \def \bsnm#1{#1} \fi
\ifx \bsuffix \undefined \def \bsuffix#1{#1} \fi
\ifx \bparticle \undefined \def \bparticle#1{#1} \fi
\ifx \barticle \undefined \def \barticle#1{#1} \fi
\ifx \bconfdate \undefined \def \bconfdate #1{#1} \fi
\ifx \botherref \undefined \def \botherref #1{#1} \fi
\ifx \url \undefined \def \url#1{\textsf{#1}} \fi
\ifx \bchapter \undefined \def \bchapter#1{#1} \fi
\ifx \bbook \undefined \def \bbook#1{#1} \fi
\ifx \bcomment \undefined \def \bcomment#1{#1} \fi
\ifx \oauthor \undefined \def \oauthor#1{#1} \fi
\ifx \citeauthoryear \undefined \def \citeauthoryear#1{#1} \fi
\ifx \endbibitem  \undefined \def \endbibitem {}\fi
\ifx \bconflocation  \undefined \def \bconflocation#1{#1} \fi
\ifx \arxivurl  \undefined \def \arxivurl#1{\textsf{#1}} \fi

\bibitem[\protect\citeauthoryear{{Bertin} and {Arnouts}}{1996}]{Bertin1996A&AS}
\begin{barticle}
\bauthor{\bsnm{{Bertin}}, \binits{E.}},
\bauthor{\bsnm{{Arnouts}}, \binits{S.}}:
\bjtitle{\aaps}
\bvolume{117},
\bfpage{393}
(\byear{1996})
\end{barticle}
\endbibitem

\bibitem[\protect\citeauthoryear{{Cao} et~al.}{2011}]{Caoli2011}
\begin{barticle}
\bauthor{\bsnm{{Cao}}, \binits{L.}},
\bauthor{\bsnm{{Ruan}}, \binits{P.}},
\bauthor{\bsnm{{Cai}}, \binits{H.}},
\bauthor{\bsnm{{Deng}}, \binits{J.}},
\bauthor{\bsnm{{Hu}}, \binits{J.}},
\bauthor{\bsnm{{Jiang}}, \binits{X.}},
\bauthor{\bsnm{{Liu}}, \binits{Z.}},
\bauthor{\bsnm{{Qiu}}, \binits{Y.}},
\bauthor{\bsnm{{Wang}}, \binits{J.}},
\bauthor{\bsnm{{Wang}}, \binits{S.}},
\bauthor{\bsnm{{Yang}}, \binits{J.}},
\bauthor{\bsnm{{Zhao}}, \binits{F.}},
\bauthor{\bsnm{{Wei}}, \binits{J.}}:
\bjtitle{Science China Physics, Mechanics, and Astronomy}
\bvolume{54},
\bfpage{558}
(\byear{2011})
\end{barticle}
\endbibitem

\bibitem[\protect\citeauthoryear{{Castelli} and
  {Kurucz}}{2004}]{Kurucz2004astroph}
\begin{botherref}
\oauthor{\bsnm{{Castelli}}, \binits{F.}},
\oauthor{\bsnm{{Kurucz}}, \binits{R.L.}}:
ArXiv Astrophysics e-prints
(2004).
\arxivurl{astro-ph/0405087}
\end{botherref}
\endbibitem

\bibitem[\protect\citeauthoryear{{ESA}}{1997}]{Tycho1997}
\begin{bbook}
\beditor{\bsnm{{ESA}}} (ed.):
\bbtitle{{The Hipparcos and Tycho Catalogues. Astrometric and Photometric Star
  Catalogues Derived from the Esa Hipparcos Space Astrometry Mission}}.
\bsertitle{ESA Special Publication},
vol. \bseriesno{1200}
(\byear{1997})
\end{bbook}
\endbibitem

\bibitem[\protect\citeauthoryear{{Fukugita} et~al.}{1996}]{Fukugita1996AJ}
\begin{barticle}
\bauthor{\bsnm{{Fukugita}}, \binits{M.}},
\bauthor{\bsnm{{Ichikawa}}, \binits{T.}},
\bauthor{\bsnm{{Gunn}}, \binits{J.E.}},
\bauthor{\bsnm{{Doi}}, \binits{M.}},
\bauthor{\bsnm{{Shimasaku}}, \binits{K.}},
\bauthor{\bsnm{{Schneider}}, \binits{D.P.}}:
\bjtitle{\aj}
\bvolume{111},
\bfpage{1748}
(\byear{1996})
\end{barticle}
\endbibitem

\bibitem[\protect\citeauthoryear{{H{\o}g} et~al.}{2000}]{Tycho2AA}
\begin{barticle}
\bauthor{\bsnm{{H{\o}g}}, \binits{E.}},
\bauthor{\bsnm{{Fabricius}}, \binits{C.}},
\bauthor{\bsnm{{Makarov}}, \binits{V.V.}},
\bauthor{\bsnm{{Urban}}, \binits{S.}},
\bauthor{\bsnm{{Corbin}}, \binits{T.}},
\bauthor{\bsnm{{Wycoff}}, \binits{G.}},
\bauthor{\bsnm{{Bastian}}, \binits{U.}},
\bauthor{\bsnm{{Schwekendiek}}, \binits{P.}},
\bauthor{\bsnm{{Wicenec}}, \binits{A.}}:
\bjtitle{\aap}
\bvolume{355},
\bfpage{27}
(\byear{2000})
\end{barticle}
\endbibitem

\bibitem[\protect\citeauthoryear{Hunter}{2007}]{Hunter:2007}
\begin{barticle}
\bauthor{\bsnm{Hunter}, \binits{J.D.}}:
\bjtitle{Computing In Science \& Engineering}
\bvolume{9}(\bissue{3}),
\bfpage{90}
(\byear{2007})
\end{barticle}
\endbibitem

\bibitem[\protect\citeauthoryear{{Ip} et~al.}{2014}]{Ip2014RAA}
\begin{barticle}
\bauthor{\bsnm{{Ip}}, \binits{W.-H.}},
\bauthor{\bsnm{{Yan}}, \binits{J.}},
\bauthor{\bsnm{{Li}}, \binits{C.-L.}},
\bauthor{\bsnm{{Ouyang}}, \binits{Z.-Y.}}:
\bjtitle{Research in Astronomy and Astrophysics}
\bvolume{14},
\bfpage{1511}
(\byear{2014})
\end{barticle}
\endbibitem

\bibitem[\protect\citeauthoryear{{Morrissey} et~al.}{2007}]{Morrissey2007ApJS}
\begin{barticle}
\bauthor{\bsnm{{Morrissey}}, \binits{P.}},
\bauthor{\bsnm{{Conrow}}, \binits{T.}},
\bauthor{\bsnm{{Barlow}}, \binits{T.A.}},
\bauthor{\bsnm{{Small}}, \binits{T.}},
\bauthor{\bsnm{{Seibert}}, \binits{M.}},
\bauthor{\bsnm{{Wyder}}, \binits{T.K.}},
\bauthor{\bsnm{{Budav{\'a}ri}}, \binits{T.}},
\bauthor{\bsnm{{Arnouts}}, \binits{S.}},
\bauthor{\bsnm{{Friedman}}, \binits{P.G.}},
\bauthor{\bsnm{{Forster}}, \binits{K.}},
\bauthor{\bsnm{{Martin}}, \binits{D.C.}},
\bauthor{\bsnm{{Neff}}, \binits{S.G.}},
\bauthor{\bsnm{{Schiminovich}}, \binits{D.}},
\bauthor{\bsnm{{Bianchi}}, \binits{L.}},
\bauthor{\bsnm{{Donas}}, \binits{J.}},
\bauthor{\bsnm{{Heckman}}, \binits{T.M.}},
\bauthor{\bsnm{{Lee}}, \binits{Y.-W.}},
\bauthor{\bsnm{{Madore}}, \binits{B.F.}},
\bauthor{\bsnm{{Milliard}}, \binits{B.}},
\bauthor{\bsnm{{Rich}}, \binits{R.M.}},
\bauthor{\bsnm{{Szalay}}, \binits{A.S.}},
\bauthor{\bsnm{{Welsh}}, \binits{B.Y.}},
\bauthor{\bsnm{{Yi}}, \binits{S.K.}}:
\bjtitle{\apjs}
\bvolume{173},
\bfpage{682}
(\byear{2007})
\end{barticle}
\endbibitem

\bibitem[\protect\citeauthoryear{{Oke} and {Gunn}}{1983}]{Oke1983}
\begin{barticle}
\bauthor{\bsnm{{Oke}}, \binits{J.B.}},
\bauthor{\bsnm{{Gunn}}, \binits{J.E.}}:
\bjtitle{\apj}
\bvolume{266},
\bfpage{713}
(\byear{1983})
\end{barticle}
\endbibitem

\bibitem[\protect\citeauthoryear{{Tan} et~al.}{2014}]{Tan2014RAA}
\begin{barticle}
\bauthor{\bsnm{{Tan}}, \binits{X.}},
\bauthor{\bsnm{{Liu}}, \binits{J.-J.}},
\bauthor{\bsnm{{Li}}, \binits{C.-L.}},
\bauthor{\bsnm{{Feng}}, \binits{J.-Q.}},
\bauthor{\bsnm{{Ren}}, \binits{X.}},
\bauthor{\bsnm{{Wang}}, \binits{F.-F.}},
\bauthor{\bsnm{{Yan}}, \binits{W.}},
\bauthor{\bsnm{{Zuo}}, \binits{W.}},
\bauthor{\bsnm{{Wang}}, \binits{X.-Q.}},
\bauthor{\bsnm{{Zhang}}, \binits{Z.-B.}}:
\bjtitle{Research in Astronomy and Astrophysics}
\bvolume{14},
\bfpage{1682}
(\byear{2014})
\end{barticle}
\endbibitem

\bibitem[\protect\citeauthoryear{{Taylor}}{2005}]{Topcat2005}
\begin{bchapter}
\bauthor{\bsnm{{Taylor}}, \binits{M.B.}}:
In: \beditor{\bsnm{{Shopbell}}, \binits{P.}},
\beditor{\bsnm{{Britton}}, \binits{M.}},
\beditor{\bsnm{{Ebert}}, \binits{R.}} (eds.)
\bbtitle{Astronomical Data Analysis Software and Systems XIV}.
\bsertitle{Astronomical Society of the Pacific Conference Series},
vol. \bseriesno{347},
p. \bfpage{29}
(\byear{2005})
\end{bchapter}
\endbibitem

\bibitem[\protect\citeauthoryear{{Wang} et~al.}{2015}]{Wangjing2015cali}
\begin{barticle}
\bauthor{\bsnm{{Wang}}, \binits{J.}},
\bauthor{\bsnm{{Cao}}, \binits{L.}},
\bauthor{\bsnm{{Meng}}, \binits{X.-M.}},
\bauthor{\bsnm{{Cai}}, \binits{H.-B.}},
\bauthor{\bsnm{{Deng}}, \binits{J.-S.}},
\bauthor{\bsnm{{Han}}, \binits{X.-H.}},
\bauthor{\bsnm{{Qiu}}, \binits{Y.-L.}},
\bauthor{\bsnm{{Wang}}, \binits{F.}},
\bauthor{\bsnm{{Wang}}, \binits{S.}},
\bauthor{\bsnm{{Wen}}, \binits{W.-B.}},
\bauthor{\bsnm{{Wu}}, \binits{C.}},
\bauthor{\bsnm{{Wei}}, \binits{J.-Y.}},
\bauthor{\bsnm{{Hu}}, \binits{J.-Y.}}:
\bjtitle{Research in Astronomy and Astrophysics}
\bvolume{15},
\bfpage{1068}
(\byear{2015})
\end{barticle}
\endbibitem

\end{thebibliography}

\end{document}